\documentclass[a4paper,fleqn,usenatbib]{mnras}
\usepackage{newtxtext,newtxmath}
\usepackage[T1]{fontenc}
\usepackage{ae,aecompl}
\usepackage{graphics}
\usepackage{graphicx}
\usepackage{amsmath}	
\usepackage[flushleft]{threeparttable}
\usepackage{booktabs,caption}

\title[EXO 2030+375]{Spectral \& Timing analysis of Be/X-ray binary EXO 2030+375 during its giant 2021 outburst}

\author[Ruchi et. al.]{
Ruchi Tamang,$^{1}$\thanks{ruchitamang76@gmail.com}
Manoj Ghising,$^{1}$\thanks{E-mail:manojghising26@gmail.com}
Mohammed Tobrej,$^{1}$\thanks{tabrez.md565@gmail.com} 
Binay Rai,$^{1}$\thanks{binayrai21@gmail.com}
\newauthor
Bikash Chandra Paul$^{1}$\thanks{bcpaul@nbu.ac.in}
\\
$^{1}$Department of Physics, North Bengal University,Siliguri, Darjeeling, WB, 734013, India
\\
}

\pubyear{2021}

\begin{document}
\label{firstpage}
\pagerange{\pageref{firstpage}--\pageref{lastpage}}
\maketitle

\begin{abstract}
 We report the X-ray spectral and timing analysis of the high mass X-ray binary EXO 2030+375  during  the 2021  type II outburst. We have incorporated NuSTAR, NICER, \textit{Swift}/BAT \& \textit{Fermi}/GBM observations to carry out a comprehensive analysis of the source. Pulse profiles in different energy ranges and time intervals have been generated and analyzed. We have performed a brief comparison of the observations amidst the peak outburst condition and also during the decaying state of the outburst.  Pulse profiles are found to evolve with time and energy.  An iron emission line at (6–7) keV is observed    in the  X-ray continuum. Distinct absorption features were observed in the spectra corresponding to the peak outburst state while such features were not detected during the later decaying phase of the outburst. We have estimated the characteristic spin-up time scale to be $\backsim$ 60 years. The continuum flux of the system and the varying luminosities covering the entire outburst period have been used to interpret the characteristics of the source. We have summarized the variability of various parameters along with their underlying physical implications . 

\end{abstract}


%
%
\section{Introduction}
EXO 2030+375 is a transient, accretion powered X-ray pulsar and is one of the member of the Be/X-ray binaries. Be/X-ray binaries are binary systems formed between a neutron star and a Be star. Neutron stars in Be/X-ray binaries (BeXRBs) revolve in a wide and moderate eccentric orbits \citep{1}.  Strong  X-ray outbursts are observed  due to an instantaneous mass accretion from the circumstellar envelope onto the neutron star while passing close to the periastron. The intensity of such an outburst rises up to a certain level than the quiescent phase.  Normally, BeXRBs  undergo periodic (Type I) X-ray outbursts observed at the periastron passage of the neutron star.  Such outbursts cover   <(20–30)\% of the orbit lasting for a few days to weeks \citep{2}. On the other hand,  giant outbursts (Type II) are also observed  from the neutron stars in BeXRBs.  Such outbursts cover an integral fraction or multiple orbits and are normally known to last for several weeks to months \citep{3,4,5,6,7}. Type II X-ray outbursts are quite rare and hence very rarely detected.  These are independent of the orbital phase or periastron passage of the binary.  The X-ray activities  in BeXRBs are  characterized by the Type-I outbursts with peak luminosity of the order of $L_{x} \leq 10^{37} erg \; s^{-1}$ and rare giant (Type-II) X-ray outbursts with peak luminosity of $L_{x} \geq 10^{37} erg \; s^{-1}$.  Depending upon the evolution of the donor, mass transfer from the companion star to the compact object occurs via capture of stellar wind or accretion from a large circumstellar disk around the companion star \citep{8}. The optical companion of such systems is a non-supergiant B or O type star bearing emission lines in its spectrum \citep{4}.

EXO 2030+375 is a well analyzed  BeXRB  associated with regular Type-I outbursts during almost every periastron passage. This transient accreting source was discovered in 1985 with EXOSAT during a Type-II outburst \citep{9} with $\sim$ 42 s pulsations. The  optical counter-part of EXO 2030+375 is a highly reddened B0 Ve star \citep{10} which shows infrared excess and  H$\alpha$ in emission \citep{11}. Using the relation between extinction and distance of sources in the Galactic plane, \cite {12} estimated the distance of EXO 2030+375 as 7.1 kpc. Type-I X-ray outbursts of EXO 2030+375, occurs almost at every periastron passage of its $\sim$ 46 day orbit \citep{13} and have been monitored with the X-ray instruments onboard RXTE, INTEGRAL, XMM-Newton, Suzaku and \textit{Swift}/BAT observatories. The  characteristic properties of the source \citep{12,14,15,16,17}, have been studied using the  available observational data. In June 2006, EXO 2030+375 was observed to undergo a giant (Type-II) X-ray outburst for the second time with an initial flux of 180 mCrab. This value is higher than the previous peak flux of about 50 mCrab observed during the entire life of the RXTE/ASM mission \citep{18}. This outburst was also followed by \textit{Swift}/BAT which reported that the peak flux  increased to 750 mCrab \citep{19}. Spin-up feature was reported in the source during the giant X-ray outbursts in 1985 \citep{9} and 2006 \citep{13} while spin-down episodes have been reported during low luminous outbursts in 1994-2002 \citep{12,20} and  also during faint outbursts after March 2016 \citep{21}. The spectra of EXO 2030+375 during  outbursts after  2006 have been described using various phenomenological and physical models which showed an iron emission line at $\sim$ 6.4 keV and interstellar absorption (\cite{17} and references therein). Various interesting features have been observed in the spectrum of the system.

Suzaku observations of  EXO 2030+375 during  Type-I outbursts in 2007 and 2012 did not reveal the presence of cyclotron absorption features in the X-ray spectrum. Presence of additional matter at certain pulse phases of the source was reported as the reason of prominent absorption dips in the pulse profiles (\cite{14}; \cite{15}). During the brighter Type-I outburst in 2007, \cite{14} reported the detection of  narrow emission lines (i.e Si XII, Si XIV, S XV) for the first time along with Fe K$\alpha$ and Fe XVI in the continuum spectrum.
Detailed analysis of the source  was carried out by using extensive RXTE pointed observations during many Type-I \& 2006 Type-II outbursts starting from 1995 to 2011 \citep{17}. Timing and spectral analysis were carried out in (3–30) keV luminosity range from $3.8 \times 10^{36}$ to $2.6\times10^{38} erg\;s^{-1} $, covering  the complete span of RXTE campaign. Temporal studies of more than 600 RXTE pointings verified the evolution of pulse profiles  with luminosity. A major peak and a minor peak at low luminosity transformed to a dual peaked profile with minor dips at high luminosity. This signifies that pulse profiles at a particular luminosity were identical irrespective of the X-ray outbursts character which reveals  that the emission geometry is mainly dependent upon the accretion rate. Since its discovery in 1985, the system has undergone X-ray outbursts regularly. Since 2015, the Type-I outbursts are found to possess diminishing intensity and eventually vanished from the light curve towards the end of 2015 or early 2016 \citep{22}. Type-I X-ray outburst activity began in early 2016 proceeding with much fainter peak luminosity. \cite{41} detected  pulsation at a minimum luminosity of $6.8\times10^{35} erg \; s^{-1}$ in (3-78) keV range, which is the lowest luminosity of the source with X-ray pulsations in the light curve. The pulsar was observed with \textit{Swift}/XRT at a fainter phase and the  data quality was found to be poor for pulsation search. Pulsations in the light curve at lower luminosity to that during the earlier \textit{Swift}/XRT observation \citep{41}, rules out the onset of propeller regime.

In this paper, we compare and reveal the characteristic features of  EXO 2030+375 during the peak state and the decaying state of the outburst. The pulse period of the source is found to be $\sim\;41.27\;\pm\;0.004$ s. An iron emission line at (6–7) keV is observed in the X-ray continuum. Absorption features are revealed in the spectrum corresponding to the peak outburst state while such features are not prominent in the decaying state (post -peak outburst state). Furthermore, several NICER observations have been taken into consideration to explore certain features of the source during the recent giant outburst. We have tried to present a comprehensive study of the source using its temporal and spectral properties by interpreting the sequential variations of various parameters .

\section{Observations and Data Reduction}

The source EXO 2030+375 previously exhibited two giant Type II outbursts in 1985 and 2006. On July 2021, the source was reported to encounter another giant Type II outburst by MAXI/GSC observations. Constant monitoring  of the source was carried out by Monitor of All-sky X-ray Image (MAXI), the Burst Alert Telescope (BAT) on the Swift satellite  and Gamma-ray Burst Monitor (GBM) on the Fermi satellite. Throughout the bursting phase, NICER started observing the source regularly. For our analysis,  we have considered observations from Neutron Star Interior Composition Explorer (NICER), Nuclear Spectroscopic Telescope Array (NuSTAR) along with some relevant observations from \textit{Fermi}/GBM and \textit{Swift}/BAT.

\subsection{NUSTAR}
The data reduction for EXO 2030+375 has been done using HEASOFT 6.29 and CALDB v 1.0.2. NuSTAR is sensitive in the (3-79) keV energy range and is the first hard X-ray focusing telescope. It consists of two independent co-aligned grazing incident telescopes which are similar but not identical. Each of the telescopes have their own focal plane module FPMA and FPMB consisting of a solid state CdZnTe detector \citep{23}. The light curves, spectra, response matrices and effective area files are generated using NUSTARDAS software v 0.4.9. The clean event files were filtered from the unfiltered event files using  the mission specific command NUPIPELINE. A circular region of 100’’ around the source center and away from the source has been taken into consideration as the required source and background regions respectively. The estimated source and background  have been used for extracting light curves and their corresponding spectra by using the  \textit{ftool} XSELECT and  imposing the mission specific command line NUPRODUCTS. For both FPMA and FPMB data, the backgrounds are individually subtracted and their light curves are combined using tool LCMATH . 
The two NuSTAR observations ( 80701320002  and  90701336002) have been considered for performing the required  timing and spectral analysis of the system. The observation coined with  observation ID 80701320002 coincides close to the peak outburst period of the source, while the other observation (90701336002)  corresponds to the decaying phase of the giant outburst respectively. NuSTAR observation ID 80701320002 and 90701336002 are hereafter referred as Obs I and Obs II respectively.

The specifications related to the NuSTAR and NICER observations under consideration are presented in \textit{table} 1.

\begin{table*}
\begin{center}
\begin{tabular}{cllllllllc}
\hline														
Observatory 	&	Observation ID	&	Date of Observation (MJD)	&	Exposure (ks)	&	Observatory 	&	Observation ID	&	Date of Observation (MJD)	&	Exposure (ks)	\\
\hline															
NuSTAR	&	80701320002	&	59456.0389	&	32.461	&		&		&		&		\\
	&	90701336002	&	59526.1466	&	23.487	&		&		&		&		\\
	\hline

NICER	&	4201960101	&	59423.2131	&	2.114	&	NICER	&	4201960123	&	59469.0088	&	2.318	\\
	&	4201960102	&	59431.008	&	1.335	&		&	4201960129	&	59481.3288	&	2.496	\\
	&	4201960103	&	59435.0881	&	2.062	&		&	4201960130	&	59483.3313	&	2.549	\\
	&	4201960104	&	59447.7555	&	0.614	&		&	4201960131	&	59485.3322	&	2.547	\\
	&	4201960105	&	59448.8604	&	0.961	&		&	4201960132	&	59487.3361	&	2.449	\\
	&	4201960106	&	59449.2481	&	1.323	&		&	4201960133	&	59489.3348	&	2.263	\\
	&	4201960107	&	59450.8667	&	0.849	&		&	4201960134	&	59491.2719	&	2.64	\\
	&	4201960108	&	59451.1829	&	7.815	&		&	4201960135	&	59493.0138	&	2.529	\\
	&	4201960109	&	59452.0963	&	4.206	&		&	4201960137	&	59497.0168	&	2.805	\\
	&	4201960110	&	59453.3183	&	3.889	&		&	4201960150	&	59526.1587	&	15.381	\\
	&	4201960111	&	59454.4093	&	2.674	&		&	4201960151	&	59527.3846	&	3.249	\\
	&	4201960112	&	59455.1194	&	2.496	&		&	4201960152	&	59529.6428	&	2.238	\\
	&	4201960113	&	59456.5959	&	2.299	&		&	4201960153	&	59530.4166	&	1.554	\\
	&	4201960115	&	59458.0882	&	2.483	&		&	4201960154	&	59532.354	&	2.041	\\
	&	4201960116	&	59459.1791	&	2.337	&		&	4201960156	&	59535.1945	&	2.778	\\
	&	4201960118	&	59463.1863	&	4.408	&		&	4201960157	&	59536.2273	&	2.445	\\
	&	4201960119	&	59465.0039	&	2.601	&		&	4201960158	&	59538.1635	&	1.951	\\
	&	4201960121	&	59466.9951	&	3.738	&		&	4201960159	&	59540.2461	&	0.418	\\
	&		&		&		&		&	4201960162	&	59544.0572	&	2.372	\\

\hline
\end{tabular}
\caption{NuSTAR and NICER observations indicated by their observation IDs along with their respective date of observations and exposure.}
\end{center}
\end{table*}

\subsection{NICER}
NICER is an International Space Station (IIS) external payload  \citep{24} designed for the study of neutron stars through soft X-ray Timing. In our work, we have used X-ray Timing Instrument (XTI), which operates in the energy range of (0.2-12.0) keV. The standard data screening and reduction of NICER observations were carried out using NICERDAS v9 and CALDBv  xti20210720.  For each observations,\;the background estimation was done using the tool nibackgen3C50 v6 \citep{25}. The clean event files were filtered using command NICERL2. The filtered event files were then loaded to XSELECT to extract the necessary source light curve \& pha files. Response and arf files were generated using command NICERARF \& NICERRMF. From the NICER clean event files, light curves in (0.7-10.0) keV energy range with a binning of 0.01s were extracted. The spectra in the energy range (0.7-10.0) keV has been considered for fitting. This is because there will be background domination above 10 keV and also the spectra below 0.7 keV has been ignored to get rid of the calibration issue. We have considered 37 NICER observations from July 28, 2021 to November 26, 2021 that lies during the outburst phase of EXO 2030+375 in 2021.

The necessary barycentric correction has been done for both NuSTAR and NICER observations to account for the orbit correction. The required spectral fitting has been carried out using XSPEC version 12.12.0 \citep{26}. 

\subsection{\textit{Fermi}/GBM}
The Fermi Gamma-Ray Space Telescope studies the gamma-ray sources and was launched in 2008. It has a Gamma-ray Burst Monitor (GBM) which is efficient in the energy range 20 MeV to about 300 GeV \citep{27}. We have considered the publicly available pulsar data for our analysis as it provides an estimate of the source flux in the hard band. The spin frequency provided by FERMI GBM Accreting Pulsars Program (GAPP) has been used to obtain the spin frequency derivative by linearly fitting three consecutive pulse frequencies \citep{28}. The rate of change of spin frequency is given by the slope of the plot and the error associated with it corresponds to 1-sigma uncertainty in measuring the spin-up rate.

\subsection{\textit{Swift}/BAT}
 Swift has been purposely designed to study the Gamma-ray burst. It has its monitoring instrument called the Burst Alert telescope which works in searching for transient events in the energy band of (15-50) keV from 80\% of the sky \citep{29}. The daily light curves  for EXO 2030 +375 has been used as an indicator of bolometric flux during the outburst which are available in the official website, the Hard X-rays Transient Monitor \citep{30}.
\section{TIMING ANALYSIS}
\subsection{NuSTAR}
We generated the light curves corresponding to the NuSTAR data with the help of \textit{ftool} LCMATH. The approximate pulse period of the source was roughly estimated by Fast Fourier Transform using the command POWSPEC. The tool EFSEARCH helps in determining the more accurate period by folding the light curve with large number of periods around the approximate period by Chi-Square maximization technique also known as the epoch-folding technique (\cite{31}; \cite{32}). Thus, the best pulse period was precisely estimated to be $\sim 41.2\;\pm\;0.004$ s. The uncertainty in the measurement of pulsation has been taken care by using method described in \cite{44}. The light curves are folded with the corresponding estimated pulse period to obtain the  pulse profile using \textit{ftool} EFOLD in the energy range of (3-79) keV. The pulse profile obtained is highly defined and displays a main pulse preceded by various intensity minima. We extracted the light curves with a binning of 0.01s. The entire energy band was resolved into various sub-intervals as (3-7) keV, (7-12) keV, (12-18) keV, (18-24) keV, (24-30) keV, (30-35) keV and (35-79) keV. Corresponding to each interval, the energy resolved pulse profiles were generated in  order  to analyze the dependency of the shape and amplitude of pulse profiles with energy as shown in \textit{figure} 2.

\begin{figure}
\begin{center}
\includegraphics[angle=0,scale=0.35]{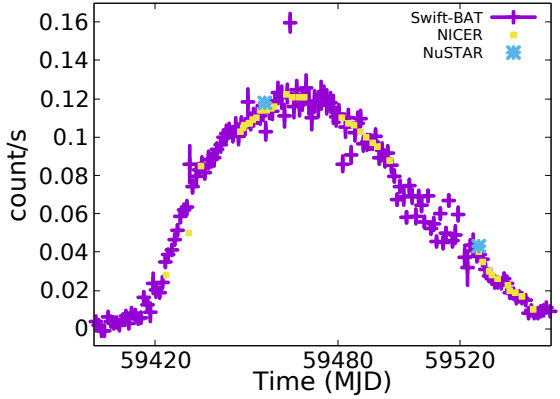}
\caption{\textit{Swift}/BAT (15-50) keV light curve of the source. The \textit{Swift}/BAT, NICER and NuSTAR observations are marked in purple, yellow and blue respectively.}
\end{center}
\end{figure}

\begin{figure}
\begin{center}
\includegraphics[angle=0,scale=0.3]{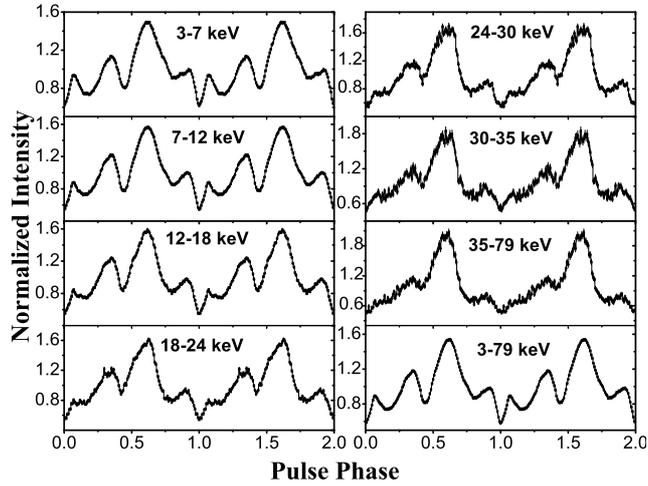}
\caption{Energy Resolved pulse profiles corresponding to NuSTAR Obs I (59456MJD) near the maximum of the outburst. Energy is expressed in keV. }
\end{center}
\end{figure}

 \begin{figure}
\begin{center}
\includegraphics[angle=0,scale=0.3]{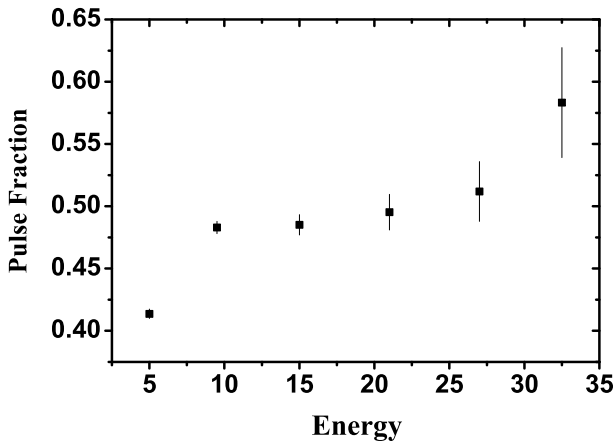}
\caption{Pulse Fraction of the source with respect to energy (keV) using NuSTAR observation Id 80701320002. The errors quoted are within 3$\sigma$ level uncertainty.}
\end{center}
\end{figure}

The pulse profile was found to evolve with time and energy. This is supported by the luminosity-resolved pulse profiles presented in \textit{figure} 5. The pulse profiles correspond to observations at different spans of time where the variability is significant. Also, the energy resolved pulse profiles presented in \textit{figure} 2 reveals distinct variabilities in the nature of the pulse profiles which characterizes the energy evolution of the pulse profiles. The variation is mostly significant in the pulse profile in the pulse phases between (0-0.2) and (0.8-1.0). As proposed by \cite{33},\; the pulse profile which are multipeaked below 20 keV evolves to single peaked nature. This feature was relevantly observed in our observations as well. The major peak was found to be dominant with increasing energy which may reflect the transition in the accretion states.  At higher energies, uncertainties are associated with the normalized count rates. The Pulse Fraction (P.F) parameter is defined as the ratio of the difference and sum of maximum $P_{max}$ and  minimum $P_{min}$ intensities of pulse profile expressed as, P.F = $\frac{P_{max}-P_{min}}{P_{max}+P_{min}}$. The variation of the pulse fraction with energy has been shown in \textit{figure} 3. The pulse fraction increases with energy by $\sim$ 7\% in the (7-12) keV range up to $\sim$ 13.4\% in the (24-30) keV energy range.

We also examined the Hardness Ratio (HR) with respect to the pulse phase of the system. The hardness ratio (HR) in the (7-12 keV \& 12-18 keV) energy band of NuSTAR observation has been illustrated in the \textit{figure} 4. The variations as observed in the HR shows spin phases where hard X-rays exceeds the soft X-rays between (0.4-0.5). Interestingly, the minima of HR lies  between pulse phase $\sim$ (0.7-0.8). Thus, we observed a phase shift in the minima of HR as compared to the minima of the pulse profile.

\begin{figure}
\begin{center}
\includegraphics[angle=270,scale=0.35]{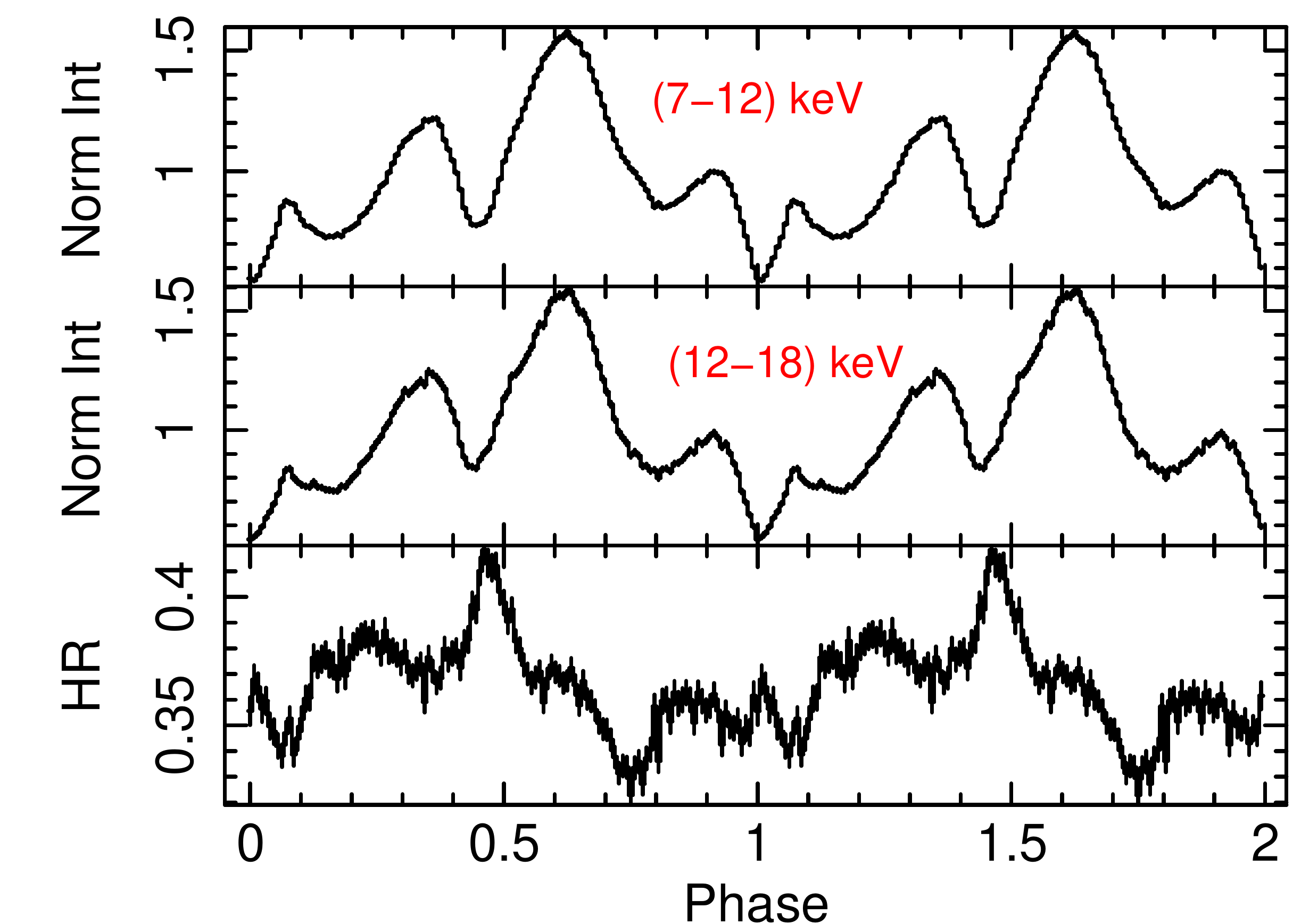}
\caption{Variation of hardness ratio with pulse phase.}
\end{center}
\end{figure}

\subsection{NICER}
For timing analysis, we considered 37 NICER observations which revealed the evolution of pulse period with time and luminosity in the soft energy band. The NICER observations in the time span  59423 MJD to  59544 MJD have been taken into consideration. Corresponding to each observation,  we estimated the approximate pulse period  and hence obtained the best period. The light curves obtained were folded  corresponding to the estimated best period to generate the pulse profiles. The pulse profiles have been observed to remarkably vary with luminosity. As the luminosity increases, the dominant X-ray beaming pattern is found to evolve from pencil-beam to a fan-beam nature on approaching close to maximum of the outburst \citep{9} with the main peak becoming more dominant as seen in the \textit{figure} 5.

\begin{figure}
\begin{center}
\includegraphics[angle=0,scale=0.3]{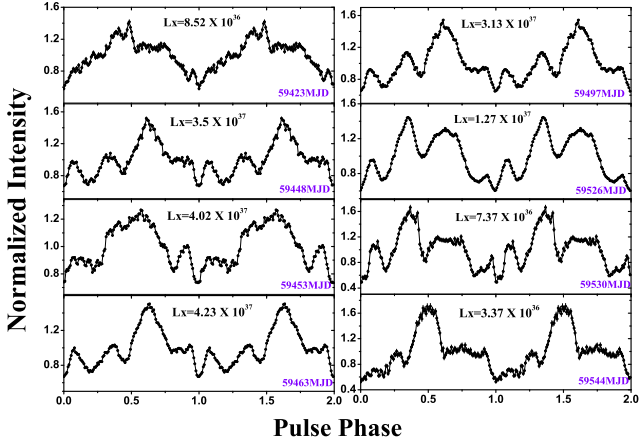}
\caption{Luminosity ($erg\;s^{-1}$) resolved pulse profiles corresponding to NICER observations of EXO 2030+375. Luminosity dependence of the pulse profiles can be clearly seen. Two pulses in each panel are shown for clarity.}
\end{center}
\end{figure}

\begin{figure}
\begin{center}
\includegraphics[angle=0,scale=0.3]{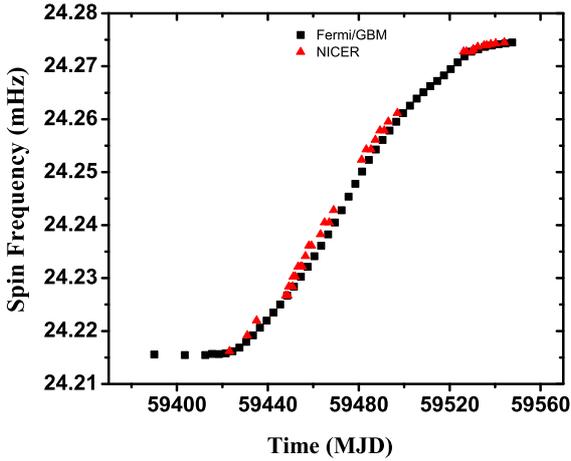}
\caption{Evolution of spin frequency with time of EXO 2030+375 during 2021 outburst. Spin frequencies estimated by \textit{Fermi}/GBM are marked in black whereas the red marks correspond to the spin frequencies obtained from NICER observations.}
\end{center}
\end{figure}

The \textit{Fermi}/GBM estimated data shows that the spin frequency has increased to $\sim$ 24.27 mHz on  59536 MJD in (15-50) keV energy range. We plotted the \textit{Fermi}/GBM estimated data along with the spin frequencies obtained for NICER observations as shown in \textit{figure} 6. 
The \textit{Swift}/BAT light curves of 2021 outburst along with the NICER and NuSTAR observations taken during different stages of the outburst have been presented alongside in \textit{figure} 1. We obtained the bolometric flux by multiplying the observed \textit{Swift}/BAT count rate by the conversion factor A$\sim\;1.072\;\times\;10^{-7}$ \citep{34}. The corresponding luminosities were computed by considering a distance of 7.1 kpc \citep{12}. 

\begin{figure}
\begin{center}
\includegraphics[angle=0,scale=0.3]{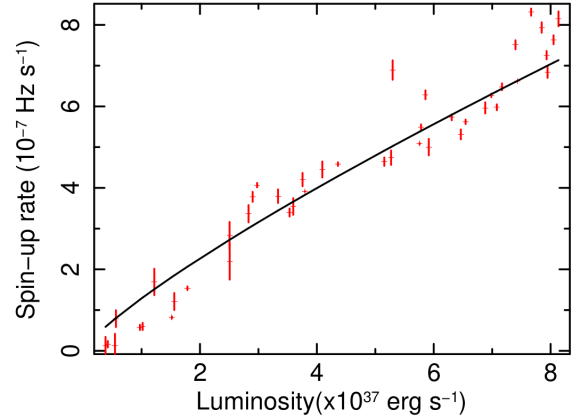}
\caption{Variation of spin-up rate with the luminosity. The solid line represents the best fitted line by using POWER-LAW model. }
\end{center}
\end{figure}

From the spin frequency data provided by \textit{Fermi}/GBM, we estimated the spin frequency derivative. The variation of spin frequency derivate with respect to luminosity has been presented in \textit{figure} 7. The frequency derivative bears a positive correlation with the luminosity of the system. The data points  are well fitted using the POWER-LAW model which is in accordance with the accretion torque model established by \cite{35} given by $\dot{\upsilon} \propto L^{6/7}$. We obtained the POWER-LAW index as 0.82 $\pm\;0.004$.

As mentioned in the preceding section, the pulse period of the source is estimated to be $\sim 41.2\;\pm\;0.004$ s. Using the $\dot{\nu}$  vs luminosity plot, we estimated the pulse period derivative.  The characteristic spin-up time scale  (-P/$\dot{P}$) is found out to be $\sim$ 60 years which is longer than those estimated during first $\sim30$ years \citep{9} and second giant outburst $\sim40$ years \citep{36}.

\section{SPECTRAL ANALYSIS} 

As analyzed earlier by \cite{14}, \cite{16}, the source EXO 2030+375 shows some complex absorption features in its spectrum and cannot be modeled using simple power-law/cutoff model. In the course of choosing  the right model for fitting the spectrum, \cite{13} used power-law/cutoff model modified with Gaussian absorption line at $\sim$ 10 keV which they interpreted as the presence of cyclotron line. However, \cite{36} fitted the spectrum considering a broad emission “bump” at $\sim$ 15 keV.   For our analysis, we first considered the NuSTAR Obs I that falls around the maximum of the outburst for spectral fitting as shown in \textit{figure} 8. The model combination of CONSTANT*PHABS*CUTOFFPL showed an emission like feature in the energy range (6-7) keV and the spectra was well fitted with the addition of GAUSSIAN model to it. This indicated the presence of He-like Fe line at $\sim$ 6.6 keV. On using the continuum model combination CONSTANT*PHABS*(CUTOFFPL+GAUSSIAN), the spectrum showed a highly significant negative residuals at $\sim$ 10 keV. We used different such model combinations to fit the broad band spectrum of the source. A CONSTANT model has been used in order to take into account for the instrumental uncertainties. Including GAUSSIAN absorption model (GABS) to the model combination and referring it as model I i.e, CONSTANT*PHABS*(CUTOFFPL+GAUSSIAN)*GABS fitted the spectrum well as it flattens the residuals  showing GABS line at $\sim$ 10.12 keV  reducing chi square from  1.61 to 1.18 .
The model combination II i.e., CONSTANT*PHABS*(CUTOFFPL+GAUSSIAN+GAUSSIAN) considering broad emission feature at $\sim$ 16 keV \citep{36} resulted in a reduced chi square value as 1.28. \cite{13} and \cite{37} also successfully used HIGHECUT model to fit the source spectra. Thus, using the continuum model III i.e.,
CONSTANT*PHABS*(HIGHECUT*POWERLAW+GAUSSIAN)*GABS *GABS reduced chi square to 1.18. The spectrum is also fitted well incorporating the Fermi-Dirac cutoff power law  (FDCUT) \citep{38} model into the model combination IV as CONSTANT*PHABS*(FDCUT+GAUSSIAN)*GABS reducing chi square value to 1.25 with GABS line energy at $\sim$ 10.33 keV. We also tried spectral fitting including the Negative and Positive power law EXponential (NPEX) \citep{39} model into the model combination V i.e, CONSTANT*PHABS*(NPEX+GAUSSIAN)*GABS which reduced chi square to 1.19 with GABS line energy at $\sim$ 10.26 keV. While using NPEX model, we kept the positive index fixed at 2 whereas the negative index was allowed to freely vary. Amongst all of the model combinations we considered, model I and model III has been found to noticeably reduce chi square value by a significant amount as compared to the other model combinations. The fitted spectra has been shown in \textit{figure} 8 and the corresponding statistical data has been presented in \textit{table} 2.
However, such features were not observed while fitting the spectrum of NuSTAR Obs II. This may be due to its comparatively low luminosity than the maximum of the outburst as it lies in the decaying phase of the outburst (59526 MJD).
We calculated the flux in (3-79) keV energy range with the help of command flux in the terminal and thus calculated the corresponding luminosity to be $\sim$ $9.49\;\times\;10^{37}\;erg\;s^{-1 }$ and $\sim$ $3.41\;\times\;10^{37}\; erg\;s^{-1}$ for Obs I and II respectively at a distance of 7.1 kpc.

\begin{table*}
\begin{center}
\begin{tabular}{cllllllc}
\hline
Parameters &	Model I		&	Model II		&	Model III		&	Model IV		&	Model V		\\
\hline
$n_{H}(cm^{-2})$	&	0.42	$\pm$0.034	&	0.83	$\pm$0.044	&	0.99	$\pm$0.044	&	1	$\pm$0.031	&	0.61	$\pm$0.038	\\
photon index	&	1.06	$\pm$0.003	&	1.19	$\pm$0.008	&	1.4	$\pm$0.007	&	1.29	$\pm$0.003	&	--	--	\\
$E_{\textit{cut}}$ (keV)	&	17.6	$\pm$0.079	&	20.11	$\pm$0.184	& 12.57	$\pm$0.098	&	5(fixed)		&	78.51	$\pm$1.135	\\
$E_{\textit{fold}}$(keV)	&	--	--	&	--	--	&	20.76	$\pm$0.229	&	16.54	$\pm$0.057	&	--	--	\\
$E_{Fe}$\;(keV)	&	6.51	$\pm$0.006	&	6.49	$\pm$0.006	&	6.51	$\pm$0.006	&	6.51	$\pm$0.006	&	6.51	$\pm$0.006	\\
$\sigma_{Fe}$\;(keV)	&	0.29	$\pm$0.008	&	0.34	$\pm$0.007	&	0.3	$\pm$0.007	&	0.31	$\pm$0.007	&	0.29	$\pm$0.007	\\
Ga \;(keV)	&	--	--	&	16.09	$\pm$0.558	&	--	--	&	--	--	&-- --	\\
Ga sigma\;(keV)	&	--	--	&	6.33	$\pm$0.429	&	--	--	&	--	--	&	--	--	\\
$E_{gabs_1}$ \;(keV)	&	10.12	$\pm$0.059	&	--	--	&	11.85	$\pm$0.148	&	10.33	$\pm$0.07	&	10.26	$\pm$0.06	\\
$E_{\sigma_1}$\;(keV)	&	2.13	$\pm$0.092	&	--	--	&	2.68	$\pm$0.24	&	1.13	$\pm$0.08	&	1.96	$\pm$0.1	\\
$Gabs_{strength_1}$ &	0.24	$\pm$0.014	&	--	--	&	0.82	$\pm$0.292	&	0.07	$\pm$0.006	&	0.19	$\pm$0.015	\\
$E_{gabs_2}$ 	\;(keV)&	--	--	&	--	--	&	17.45	$\pm$2.638	&	--	--	&	--	--	\\
$E_{\sigma_2}$ \;(keV)	&	--	--	&	--	--	&	4.89	$\pm$1.769	&	--	--	&	--	--	\\
 $Gabs_{strength_2}$	&	--	--	&	--	--	&	0.68	$\pm$0.437	&	--	--	&	--	--	\\
	&			&			&			&			&			\\
	&			&			&			&			&			\\
$\chi_{\upsilon}^{2}$	&	3272.05		&	3560.36		&	3265.59		&	3456.22		&	3311.17		\\
d.o.f	&	2776		&	2776		&	2772		&	2776		&	2775		\\
	&			&			&			&			&			\\
Reduced $\chi_{\upsilon}^{2}$	&	1.18		&	1.28		&	1.18		&	1.25		&	1.19		\\

\hline
\end{tabular}
\caption{Best-fit model parameters obatained from the spectral fitting of NuSTAR Obs I of EXO 2030+375 during 2021 outburst with Model I  \textsc{(CONSTANT*PHABS*(CUTOFFPL+GAUSSIAN)*GABS)}, Model II   \textsc{(CONSTANT*PHABS*(CUTOFFPL+GAUSSIAN+GAUSSIAN))}, Model III  \textsc{(CONSTANT*PHABS*(HIGHECUT*POWERLAW+GAUSSIAN)*GABS*GABS)}, Model IV \textsc{(CONSTANT*PHABS*(FDCUT+GAUSSIAN)*GABS)} and Model V \textsc{(CONSTANT*PHABS*(NPEX+GAUSSIAN)*GABS)}.The hydrogen column density $n_{H}$ is of the order of $10^{22}$ $cm^{-2}$. Errors are within 90\% confidence interval. Ga in the table denotes the second \textsc{Gaussian model} used in the Model II.}.
\end{center}
\end{table*}

\begin{figure}
\begin{center}
\includegraphics[angle=0,scale=0.4]{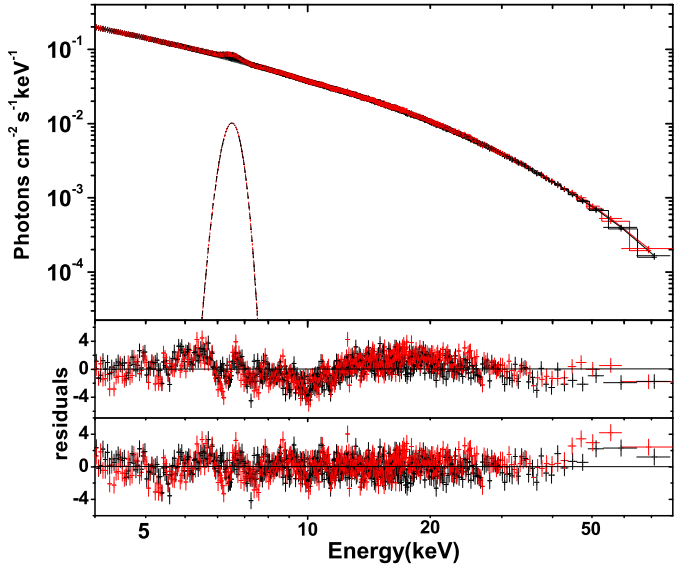}
\caption{Broad band spectrum of EXO 2030+375 using NuSTAR observation. The top panel represents the unfolded spectrum. Negative residuals can be observed at $\sim$ 10 keV as relevant from the middle panel.The bottom panel represents the best fitted spectra.}
\end{center}
\end{figure}

For all the NICER observations taken under consideration for analysis, we obtained their respective fluxes and the corresponding luminosities. Observations made throughout the outburst, revealed the enhancement of the continuum flux from $1.41\;\times\;10^{-9}\;erg\;cm^{-2}\;s^{-1}$ on July 28, 2021 and reached its maximum value of  $7.02\times10^{-9}\;erg\;cm^{-2}\;s^{-1}$  on September 6th, 2021. After attaining its maximum value, the (0.7-10.0) keV flux gradually decreased with time to a minimum of $\sim\; 5.58\;\times\;10^{-10}\; erg\;cm^{-2}\;s^{-1}$ on around 59544 MJD. The variation of luminosity with respect to time has been shown in \textit{figure} 9. It can be clearly seen that the maximum luminosity attained by the source is $\sim\;4.23\;\times\;10^{37} \;erg\;s^{-1}$ which gradually decreases  to its minimum  value of $\sim3.37\times10^{36}\;erg\;s^{-1}$ on 59544 MJD. The spectras have been fitted in the energy range of (0.7-10.0) keV using a single continuum model PHABS*(BBODY+POWERLAW). For the observations with luminosity of the order $\sim10^{37}\;erg\;s^{-1}$, we used an additional GAUSSIAN model to the continuum model. The emission like feature observed between (6-7) keV was well fitted by GAUSSIAN model  reducing chi square from $\sim$1.98 to 1.49 for a particular NICER observation. However, for the observations having luminosity $\leq10^{37} \;erg\;s^{-1}$, no such emission feature was observed.

\begin{figure}
\begin{center}
\includegraphics[angle=0,scale=0.3]{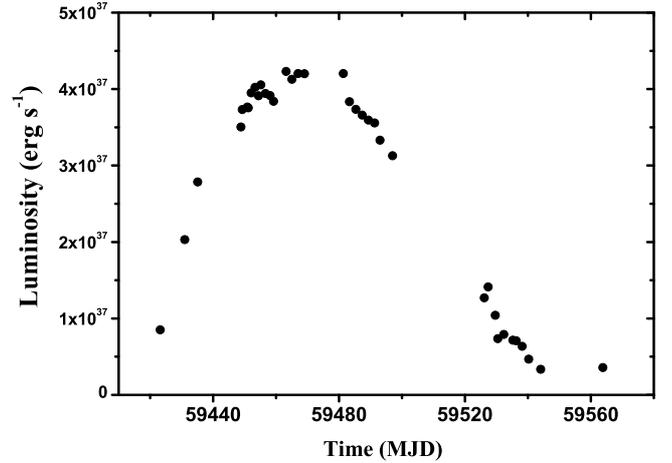}
\caption{Variation of Luminosity ($erg \; s^{-1}$) with respect to time as obtained from NICER observations.}
\end{center}
\end{figure}

\begin{figure}
\begin{center}
\includegraphics[angle=0,scale=0.3]{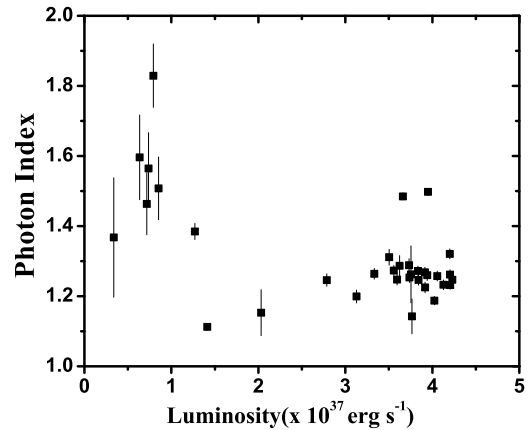}
\caption{Variation of power-law photon index with 0.7-10.0 keV luminosity \;($\times 10^{37} erg\;s^{-1}$) \; are shown for NICER observations of EXO 2030+375. Luminosities are measured assuming the distance of the source as 7.1 kpc.}
\end{center}
\end{figure}

From the spectral analysis of these NICER observations, we observed the dependency of photon index with luminosity. Thus, we plotted the photon index for all the NICER observations along with the observed luminosity as shown in \textit{figure} 10. A bi-modal behavior in the variation of the power-law index with increasing luminosity was observed. This observation validates the fact that the spectral shape depends on the rate of mass accretion. A negative correlation between the power-law photon index and luminosity signifies the pulsar spectrum to be relatively soft at lower luminosities. However, a positive correlation between power-law photon index and luminosity can be observed as luminosity approaches higher values. The pulse fraction shows negative correlation with luminosity as can be seen in \textit{figure} 11.

\begin{figure}
\begin{center}
\includegraphics[angle=0,scale=0.3]{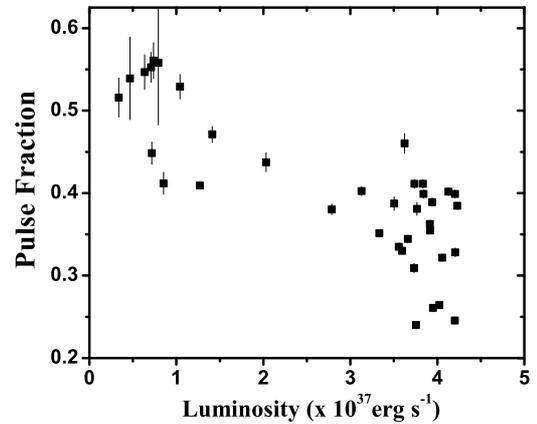}
\caption{Variation of pulse fraction with  luminosity in (0.7-10) keV energy range corresponding to NICER observations.}
\end{center}
\end{figure}

\section{PHASE RESOLVED SPECTROSCOPY}

The phase resolved spectral analysis of the NuSTAR Obs Id 80701320002 was performed in order to understand the variation of different spectral parameters with the pulse phase. Firstly, we divided each pulse into ten equal intervals and their respective good time interval (GTI) files were created. These GTI files were then used for the generation of FPMA and FPMB spectras using the
tool XSELECT. The spectras obtained for different phase intervals are fitted again with the same continuum model consisting of CONSTANT, PHABS, CUTOFFPL, GASUSSIAN and GABS. The flux and corresponding luminosities are calculated for each intervals .We observed that with the pulse phase, the corresponding flux value varies between ($1.54-1.57) \;\times\;10^{-8} \;erg\;cm^{-2}\;s^{-1}$. The photon index value ($\alpha$) and the highecut energy ($E_{cut}$) varies between (1.08-1.11) and (17.05-17.94) keV respectively. The  value of Fe emission line initially decreases from 6.55 to 6.48 keV and then again suddenly increases from 6.52 to 6.54 keV between phase (0.5-0.8). However, an abrupt decrease in its value to 6.50 keV is observed between phase (0.8-0.9) after which again it gradually increases to its maximum of 6.55 keV. The maximum value of equivalent width of He like iron ($\sigma_{Fe}$)  has been obtained at 0.38 keV which is observed between (0.4-0.5). The absorption feature $E_{gabs}$ has its maximum at 10.27 keV in  between interval (0-0.1) and then slightly decreases reaching a value of 8.4 keV in between (0.1-0.2). It again increases to 9.1 keV in between (0.3-0.4) followed by decrease in its value towards minimum at 7.98 keV between interval (0.8-0.9) and then rises again to 9.67 keV in (0.9-1.0) interval. The maximum value of $\sigma_{gabs}$ is 4.53 keV which lies between phase interval (0.1-0.2). The gabs strength value varies between (0.23-0.92) keV attaining its minimum in (0-0.1) and maximum in (0.1-0.2) phase interval. From \textit{figure} 12, we can see that the  width ($\sigma_{gabs}$) and strength of Gabs line shows negative correlation with Gabs line energy ($E_{gabs}$). The absorption feature at $\sim$ 10 keV has been observed in all of the phases.

\begin{figure}
\begin{center}
\includegraphics[angle=0,scale=0.3]{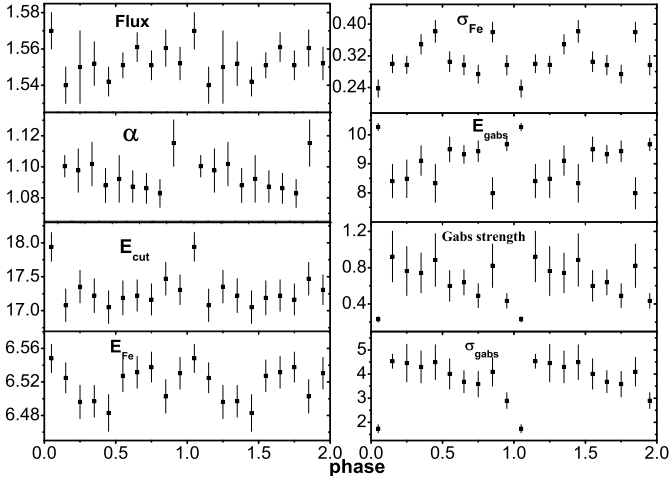}
\caption{Modulation of phase-resolved parameters using Obs I of NuSTAR represented by ID 80701320002  observed on 59456.0389 MJD. Flux is in the order of $10^{-8}$ erg\;$cm^{-2}$\;$s^{-1}$. $E_{cut}$, $E_{Fe}$ and $E_{gabs}$ are measured in keV.}
\end{center}
\end{figure}

\section*{Discussion}

In this paper, we have presented the timing and spectral analysis of EXO 2030+375 using two NuSTAR observations, one lying on the peak and the other along the decaying phase of the outburst. In addition to it, we considered 37 NICER observations almost covering the entire period during its 2021 giant (type II) outburst. The source luminosity has been observed to be as high as $\sim$ $9.49\;\times\;10^{37}\;erg\;s^{-1 }$ in (3.0-79.0) keV energy range with NuSTAR observation taken on 59456 MJD. However, in the soft energy band (0.7-10.0 keV), the maximum luminosity of the source has been observed to be  $\sim$ $4.23 \;\times\;10^{37} erg\;s^{-1}$ by considering NICER observations. The pulse profile has been observed to evolve with luminosity revealing a transition from pencil-beam geometry to fan-beam nature on approaching maximum luminosity limit. The luminosity-resolved pulse profiles correspond to observations at different points of time. Therefore, the variability in the nature of the pulse profiles represented in \textit{figure} 5 reveals the time evolution of pulse profiles. We found the spinning up of the pulsar during the outburst until it reached an equilibrium phase. In case of an accreting neutron star, the pulse period variation is related to the interaction between magnetosphere of the neutron star and the accreting gas. According to the standard theory, the condition for accretion is that the radius of the magnetosphere of the neutron star should be lesser than the corotation radius \citep{40}. The matter falls into the stellar surface if the rotational velocity of the accreting matter is less than the Keplerian velocity. However, if the magnetospheric radius is greater than or equal to the corotation radius, the centrifugal barrier emerges and the matter is thrown out beyond the capture radius by the magnetic field. The neutron star then enters the propeller regime which results in a distinct spin down evolution in its pulse period. The energy dependence of the pulse profiles is relevant from \textit{figure} 2. The energy-resolved pulse profiles are found to attain distinct variabilities transforming from a fan-beam to a pencil-beam nature significantly observed above $\sim$ 30 keV suggesting a transition in the accretion regime of the pulsar from sub-critical to the super-critical regime. 
  
   In our study with NICER observations, when the source luminosity was as low as $\sim 3.36 \;\times\;10^{36} erg\;s^{-1}$,
   it is found to exhibit a pulse period of $\sim 41.195\;s$ on 59544 MJD. The \textit{Fermi}/GBM data shows that the source attained a maximum spin frequency i.e., minimum pulse period of $\sim$ 41.194 s on 59553 MJD. This can be interpreted as an equilibrium period after which the source has been found to experience a spin down for a short span of time until 59559 MJD. However, the \textit{Fermi}/GBM data reveals that the source started spinning up again after 59559 MJD.  Careful analysis of the archival data and previously reported informations have been used to produce substantial results. The hustle related to the negative residuals associated with the NuSTAR observations adds more complexity to it. Such prominent observations and suitably fitted spectra leads us to a temptation of quoting the feature observed  at $\sim$ 10 keV  as a cyclotron line. Further, this feature was also distinctly observed in the phase resolved spectra corresponding to all the phases adding more weightage to this fact. However, the theories predicted in the past \citep{37} have led to uncertain aspects about it. \cite{37} also considered relation between CRSF energy and spectral cutoff energy (refer fig 9 of \cite{43}) questioning the declaration of 10 keV feature as CRSF to be unusual. This 10 keV feature that has been found in many pulsars is unclear which can be due to drawback of semi phenomenological models used in spectral fitting \citep{43}. For simplicity, if we  interpret the feature $\sim$ 10.12 keV as a cyclotron line \citep{13}, then the magnetic field associated with the system  can be computed using the expression: $E_{cyc} =11.6B_{12}\times(1+z)^{-1}$ keV
where $B_{12} $ represents the magnetic field strength in units of $10^{12}$ G and z $\sim$ 0.3 is the gravitational redshift in the scattering region corresponding to standard neutron star parameters. 
Assuming a distance of 7.1 kpc, the magnetic field associated with the system was found out to be $\sim 1.13 \times 10^{12} G$. This is in accordance with the typical field of a neutron star. Interestingly, the critical luminosity of the system was found to show coincidence with the transitional luminosity observed in the pulse profile and photon index features. Amidst all the twists and turns, we have made use of NICER observations for deeper understanding of the source. Due to limitations in the feasibilty of energy ranges in NICER, we noticed the presence of an iron emission line $\sim$ 6.6 keV in the X-ray continuum observed in the (0.7 -10) keV energy range. However, no absorption features were observed in the said energy band.\;A very interesting character of the Iron emission line was observed in due course of our analysis. The width of the line was found to diminish with luminosity which dissapeared as the luminosity approached to the order of $\sim$ $10^{36} erg \;s^{-1} $ . The spectral analysis of the NICER observations, revealed the dependence of photon index with luminosity. A bi-modal behaviour in the variation of the power-law photon index with increasing luminosity is clearly observed from \textit{figure}10. At low luminosities, the power-law photon index is found to be anti-correlated with the luminosity of the source which validates the fact that the spectral shape depends on the rate of mass accretion. Negative correlation between the power-law photon index and luminosity reflects the spectrum to be relatively soft at low luminosities. However, a positive correlation between powerlaw photon index and luminosity was seen at higher luminosities. This may be a result of two different accretion regimes \citep{60} defined by a certain critical luminosity \citep{61}. In the sub-critical accretion regime, an increase in luminosity leads to an increase in the height of emission zone in accretion column to approach towards the object's surface. However, in the super-critical accretion regime,  the emission zone shifts up in the accretion column. The Comptonization of photons takes place between the radiative shock and the sinking zone of the object. This region is small in supercritical state while its height increases with increasing luminosity \citep{61}. The region between the radiative shock and the neutron star is linked with a balanced rate of diffusion and advection which lowers the velocity of the comptonising electrons. As a result, the Comptonization process is not strong enough to provide an adequate amount of energy to the photons to excite them which leads to softness of the photons with increasing luminosity. So, the positive correlation of photon index with luminosity is established. However, in the sub-critical regime,  the height of the emission region is found to deplete with increasing luminosity. This depletion in the size of the sinking region, causes an increase in the optical depth leading to an increase in hardness of photons \citep{60}. This agrees with the hardening of photons with increase in luminosity. The pulse fraction bears an overall anti-correlation with luminosity. The frequency derivative is positively correlated with the luminosity of the system. The data points fitted using the POWER-LAW model is in accordance with theoretical predictions. The source under consideration has been known to undergo frequent astronomical activities and will hopefully be explored further with more concrete evidences related to the features of the source.

 \section*{Acknowledgement}
 The publicly available data of the pulsar provided by High Energy Astrophysics
Science Archive Research Center (HEASARC) Online Service data archive has been used for our study. The \textit{Fermi}/GBM and \textit{Swift}/BAT data used in the research is provided by Fermi-GAPP and Swift/BAT team respectively. We would like to thank anonymous reviewer for his/her valuable comments and suggestions which helped us in improving the manuscript both in quality and quantity. RT would like to acknowledge CSIR/NET for a research grant 09/0285(11279)/2021-EMR-I. Authors are thankful to IUCAA Center for Astronomy Research and Development (ICARD), Physics Dept, NBU for extending research facilities.
 
 \section*{Data availability}
 
 The observational data used in this study can be accessed from the HEASARC data archive and is publicly  availaible for carrying out research work.









\bsp	
\label{lastpage}
\end{document}